\documentclass[letterpaper,showpacs,preprintnumbers,twocolumn,eqsecnum,superscriptaddress,aps,nofootinbib]{revtex4}
\usepackage{amssymb}\usepackage[centertags]{amsmath}\usepackage{txfonts}\usepackage{epsfig}\usepackage{bm}
\usepackage{color}\usepackage{graphicx,graphics}\usepackage{multirow}\usepackage{float}\usepackage{ulem}
\usepackage[dvipdfm,pdfstartview=FitH]{hyperref}\usepackage{setspace}\usepackage{slashed}\usepackage{booktabs}
\hypersetup{colorlinks=true, citecolor=blue, linkcolor=red, filecolor=black,urlcolor=blue}
\allowdisplaybreaks[4]

\begin{document}

\title{Parity odd Fragmentation Functions}

\author{Weihua Yang } 

\affiliation{Reliability and Environmental Engineering National Key Laboratory, Beijing Institute of Spacecraft Environment Engineering, Beijing 100094, China}

\begin{abstract}
    Quantum chromodynamics is a non-Abelian gauge theory of strong interactions, in which the parity symmetry can be violated by the non-trivial $\theta$-vacuum tunneling effects. The $\theta$-vacuum induces the local parity odd domians. Those reactions that occur in these domains can be affected by the tunneling effects and quantities become parity odd. In this paper we consider the fragmentation process where parity odd fragmentation functions are introduced. We present the fragmentation functions by decomposing the quark-quark correlator. Among the total 16 fragmentation functions, 8 of them are parity conserved, and the others are parity violated. They have a one-to-one correspondence. Positivity bounds of these one dimensional fragmentation functions are shown. To be explicit, we also introduce a operator definition of the parity odd correlator. According to the definition, we give a proof that the parity odd fragmentation functions are local quantities and vanish when sum over all the hadrons h.
\end{abstract}


\maketitle

\section{Introduction}\label{S:introduction}

Quantum chromodynamics (QCD) is a fundamental non-Abelian gauge theory of strong interactions. It has two outstanding properties, asymptotic freedom and quark confinement.
However, the axial vector current is not conserved in QCD because it can obtain quantum corrections from the triangle diagrams duo to the complex structure of the QCD vacuum. 
It is known that the axial vector current anomaly, Adler-Bell-Jackiw anomaly \cite{Adler:1969gk}, is related to a deviation of the QCD Lagrangian which is known as the $\theta$-term \cite{Fujikawa:1979ay,Peskin:1995ev,Crewther:1978zz}, $\frac{\theta g^2}{16\pi^2} \tilde G^{\mu\nu}G_{\mu\nu}$. The $\theta$-term indicates that the physical vacuum, $\theta$-vacuum, state is a superposition of the $n$-vacua states ~\cite{tHooft:1976rip,Jackiw:1976pf,Callan:1976je}. The vacuum to vacuum transition amplitude is determined by the topological charge (density) which is just equal to the $\theta$-term.
The parity violating effect is not expected in perturbative QCD, but the $\theta$-term in the QCD Lagrangian can violate the parity and charge-parity symmetries and give rise to the strong CP problem \cite{Peccei:2006as,Cheng:1987gp}.
Though the measurements of electric dipole moment of neutron indicate that the parity violation is $local$~\cite{Baker:2006ts}, it has been shown that the local P-odd effects can be directly observed~\cite{Kharzeev:1998kz}, e.g, the P-odd $\theta$-term can leads to the famous chiral magnetic effect in heavy ion collisions~\cite{Kharzeev:2004ey}. Even in the fragmentation process, the P-odd effects play a role of inducing the P-odd fragmentation functions (FFs) and the effects can be detected in experiments through physical observables, e.g., handedness correlation and azimuthal asymmetries~\cite{Efremov:1995ff,Kang:2010qx,Yang:2019rrn}.

Thanks to the asymptotic freedom, many high energy reactions can be studied in the form of factorization theorems \cite{Collins:1989gx}, which separate the calculable hard parts from the non-perturbative soft parts in the cross sections. If only the fragmentation process is taken into consideration, the non-perturbative soft parts are usually factorized as FFs. In field theory the FFs are often given by the operator definitions which are determined by the quark-quark correlators.
The quark-quark correlators satisfy two constraints, Hermiticity and parity conservation. Because of the presence of the gauge link and final state interactions between the specified hadron and the rest debris from the collision, time-reversal puts no constraint on the correlator. By considering the non trivial $\theta$-vacuum tunneling effects, parity conservation also imposes no constraint on the correlator. In this case the P-odd FFs can emerge as well as the P-even FFs in the fragmentation process \cite{Kang:2010qx}.

Being important quantities in describing the high energy reactions, FFs are non-perturbative quantities which cannot be calculated with perturbative theory, they are mainly determined by phenomenological models and experiment data.
In practice to determine these FFs, positivity bounds have to be used. Positivity bounds are model-independent constraints for  FFs. They can be derived by using the optical theorem which relates these FFs to parton-hadron scattering amplitudes \cite{Soffer:1994ww,Goldstein:1995ek,Metz:2016swz}. In this paper we discuss these properties of the P-odd FFs. We also separate the quark-quark correlator into a P-odd part and a P-even part and introduce a definition of the P-odd correlator. According to the definition of the P-odd correlator, we prove that FFs are local quantities which vanish when sum over all the hadrons h. This indicates that the P-odd FFs can be detected only on event-by-event basis.

In this paper we only consider FFs.  In Sect. \ref{S:thetavacuum} we give a brief introduction to the non- trivial $\theta$-vacuum which induces that P-odd FFs in the fragmentation process. The decomposition of the correlator at leading twist is given in Sect. \ref{S:functions}. In Sect. \ref{S:bounds} we present the positivity bounds of FFs. By introducing a definition of the P-odd correlator, we also present a proof that the P-odd FFs are local quantities. Finally, a brief summary is given in Sect \ref{S:summary}.

\section{Non-trivial $\theta$-vacuum }\label{S:thetavacuum}

As mentioned in the introduction, the axial vector current is not conserved in QCD, it can obtain the quantum corrections from the triangle diagrams duo to the complex structure of the QCD vacuum. The divergence of the axial vector current is given by
\begin{align}
  \partial_\mu j^{\mu5}=\frac{g^2}{16\pi^2}\tilde G^{\mu\nu}G_{\mu\nu}, \label{f:divergenceresult}
\end{align}
where $g$ is the strong interaction coupling constant. $G_{\mu\nu}$ is the full field strength tensor of the gauge field. The dual field strength tensor is  defined as $\tilde G^{\mu\nu}=\frac{1}{2}\epsilon^{\alpha\beta\mu\nu}G_{\alpha\beta}$.
In fact, the pseudoscalar term can be written as a total divergence
\begin{align}
  \tilde G^{\mu\nu}G_{\mu\nu}=\partial_\mu K^\mu,  \label{f:partialK}
\end{align}
with
  $K^\mu = \varepsilon^{\mu\alpha\beta\gamma}A_\alpha^a \big[G_{\beta\gamma}^a-\frac{g}{3}f^{abc}A_\beta^b A_\gamma^c\big]$. 
Using the boundary condition $A_\mu=0$ at spatial infinity, one finds that the axial vector current satisfies the equation,
\begin{align}
  \int d^4x \partial_\mu j^{\mu5}= \frac{g^2}{16\pi^2} \int d^4x \partial_\mu K^\mu =\frac{g^2}{16\pi^2} \int d\sigma_\mu K^\mu . \label{f:axialequation}
\end{align}
It is known that integral is zero because of the surface ($\sigma_\mu$) integral.
However, this equation is not correct because the boundary condition is not being chosen correctly. 't Hooft argued that $A_\mu$ should be a pure gauge at spatial infinity \cite{tHooft:1976rip}.
The vacuum state of a theory is often defined as the state with the minimal energy. In a non-Abelian gauge theory, the minimal energy can be defined to be zero with the configuration $A_\mu=0$. This is not the only configuration with zero energy, because every transformation of $A_\mu=0$ is still a state with minimal energy,
\begin{align}
  A_\mu\to \Omega A_\mu \Omega^{-1} + \frac{i}{g} (\partial_\mu \Omega) \Omega^{-1}. \label{f:Atrans}
\end{align}
Putting $A_\mu=0$ into this equation yield the configuration of pure gauge,
  $A_\mu^{pure} =\frac{i}{g} (\partial_\mu \Omega) \Omega^{-1}$. 
In the temporal gauge $A_0 =0$, one can classify these vacuum configuration by requiring $\Omega$ going to unity as $r\to \infty$,
\begin{align}
  \Omega\to e^{i2\pi n}, \quad r\to \infty,  \quad   n= 0, \pm1, \pm 2, \cdots. \label{f:omega}
\end{align}
This assumption ensures that the surface integral over the current $K^\mu$ in Eq. ({\ref{f:axialequation}}) does not vanish. The integer $n$ which is closely related to the current is determined by an integral over the pure gauge fields \cite{Crewther:1978zz},
\begin{align}
  n=\frac{g^2}{16\pi^2}\int d^3r K^0_{n}, \quad  K^0_{n} =-\frac{g}{3}f_{ijk}\epsilon_{abc}A^{ia}_{n}A^{jb}_{n}A^{kc}_{n}. \label{f:nwinding}
\end{align}

It is known that the physical vacuum state is a superposition of the $n$-vacua sates \cite{tHooft:1976rip,Jackiw:1976pf,Callan:1976je}.
Assuming the $n$-vacua state, labeled by the winding number $n$, is $|n \rangle$. Then the true physical vacuum, $\theta$-vacuum, state can be expressed as a superposition of $|n \rangle$, $|\theta \rangle=\sum_n e^{-in\theta} |n\rangle$, with $\theta$ being a real number.
Considering a vacuum to vacuum transition between two vacua at $t=\pm \infty$, we have
\begin{align}
   _+\langle \theta | \theta \rangle_- =\sum_\nu e^{i\nu\theta} \sum_n \langle n+\nu| n\rangle.  \label{f:thetatrans}
\end{align}
By using the path integral formation, the transition amplitude can be expressed by \cite{Crewther:1978zz},
\begin{align}
  _+\langle \theta | \theta \rangle_-=\sum_\nu \int \delta A e^{iS_{eff}[A]}\delta\Big(\nu-\frac{g^2}{16\pi^2} \int d^4x \tilde G^{\mu\nu}G_{\mu\nu}\Big), \label{f:thetapath}
\end{align}
where $\nu$ is the difference of the winding numbers and is given by
the transition from a configuration with $n_-$ at $t=-\infty$ to one with $n_+$ at $t=+\infty$,
\begin{align}
  \nu=
  \frac{g^2}{16\pi^2} \int d\sigma_\mu K^\mu\Big|^{t=+\infty}_{t=-\infty} =\frac{g^2}{16\pi^2} \int d^4x \tilde G^{\mu\nu}G_{\mu\nu}. \label{f:vwinding}
\end{align}
Introducing the topological charge density, $\partial_\mu j^{\mu5}=Q$, we obtain $\nu=\int d^4x Q$. This indicates the vacuum transition is determined by the topological charge (density).

In Eq. (\ref{f:thetapath}), the effective action $S_{eff}[A]= S_{QCD}[A]+\int d^4x \theta Q$, this means  $\theta Q=\frac{g^2\theta}{16\pi^2}\tilde G^{\mu\nu}G_{\mu\nu}$ has been added to the customary QCD Lagrangian:
\begin{align}
  \mathcal{L}=-\frac{1}{4}(G_{\mu\nu})^2+\bar\psi(i\slashed D-m)\psi + \frac{\theta g^2}{16\pi^2} \tilde G^{\mu\nu}G_{\mu\nu}. \label{f:thetalagrangian}
\end{align}
By utilizing Eq. (\ref{f:divergenceresult}) with $j^{\mu5}=\bar \psi\gamma^\mu\gamma^5\psi$, the Lagrangian can be rewritten as:
\begin{align}
  \mathcal{L} &=-\frac{1}{4}(G_{\mu\nu})^2+\bar\psi(i\slashed \partial-m)\psi + \bar\psi\gamma^\mu(gA_\mu - \tilde\theta_\mu\gamma^5)\psi , \label{f:Ldelta}
\end{align}
where $\tilde\theta_\mu =\partial_\mu \theta $. Since $\theta$ is a pseudoscalar field, $\tilde\theta$ can be taken as a pseudovector. $\tilde \theta$ is different from the vector field potential, $A_\mu$, it plays a role of the potential coupling to the axial vector current which is determined by the topological charge (density) Q.

It is can be seen that the $\theta$-term in the QCD Lagrangian is parity violated ($\tilde G^{\mu\nu}G_{\mu\nu}=-4\vec B\cdot \vec E$). It forms a domain in which interactions are affected by the non-trivial $\theta$-vacuum tunneling effects. Fragmentation functions are P-even quantities because strong interaction is parity conserved. However, P-odd FFs can emerge when fragmentation processes go through these P-odd domains. In the following context we present the FFs for spin-1/2 hadrons and shown some properties of them.

We note here that the $\theta$-term have one more origin, the chiral transformation. We do not illustrate this transformation in this paper, one can refer to the famous textbook \cite{Peskin:1995ev} for details.

\section{Fragmentation functions} \label{S:functions}

In the quantum field theoretical formulation, FFs are given by the quark-quark correlators. The quark-quark correlator takes the following form,
\begin{align}
  \hat \Xi(k,p_h)=\frac{1}{2\pi} & \int d^4\xi e^{ik\xi} \sum_X \langle 0 |\mathcal{L}(\xi,\infty)\psi(\xi)|p_h, X\rangle \nonumber\\
   &\times \langle p_h, X|\bar\psi(0)\mathcal{L}^\dag(0,\infty)|0 \rangle, \label{f:correlationfunction}
\end{align}
where $k$ and $p_h$ are the momenta of the quark and produced hadron. $\mathcal{L}(\xi,\infty)$ is the gauge link. Because of the transverse component of gauge field, the gauge link does not disappear in light-cone gauge $A^+=0$. However, the appearance of gauge links in Eq. (\ref{f:correlationfunction}) has no influence  on the following discussions. Therefore, we just omit it for simplicity in the following context.
By integrating over $k^-$, we can obtain the transverse momentum dependent (TMD) correlator which is a $4\times 4$ matrix in Dirac space depending on the hadron state. Thus, the correlator can be decomposed in terms of the $\Gamma$ matrices, i.e., $\Gamma = \{I, i\gamma^5, \gamma^\rho, \gamma^\rho\gamma^5, i\sigma^{\rho\sigma}\gamma^5$\}. The decomposition can be written explicitly as,
\begin{align}
  \hat\Xi=I\Xi+i\gamma^5\tilde\Xi +\gamma^\alpha \Xi_\alpha +\gamma^\alpha\gamma^5 \tilde\Xi_\alpha+i\sigma^{\alpha\beta}\gamma^5\Xi_{\alpha\beta}. \label{f:XiD}
\end{align}
As a consequence of parity constraint, at leading twist the coefficients in Eq.~(\ref{f:XiD}) can be rewritten as the products of Lorentz covariants and scalar functions \cite{Chen:2016moq},
\begin{align}
  &z\Xi_\alpha=\bar n_\alpha\bigg[D_1+\frac{\varepsilon_{T kS}}{M}D^\perp_{1T}\bigg],\label{f:zxi}\\
  &z\tilde\Xi_\alpha=\bar n_\alpha\bigg[\lambda G_{1L}+\frac{k_T \cdot S_T}{M}G^\perp_{1T}\bigg],\label{f:zxi5}\\
  &z\Xi_{\rho\alpha}=\bar n_\rho\bigg[\frac{\varepsilon_{T k\alpha}}{M}H^\perp_{1} +S_{T\alpha}H_{1T}+\frac{k_{T\alpha}}{M}H^\perp_{1S}\bigg].\label{f:zxiodd}
\end{align}
where $\bar n$ is lightlike unit vector, $\lambda$ and $S_T$ are the helicity and the transverse component of the spin of the nucleon. We have defined the shorthanded notation $H^\perp_{1S}=\lambda H^\perp_{1L}+\frac{k_T \cdot S_T}{M}H^\perp_{1T}$. For simplicity, we have omit the arguments $(z, k_T)$ of these function. We also use $D, G$ and $H$ to denote the unpolarized, longitudinal polarized and transverse polarized quarks distributions. The subscript $L, T$ denote the longitudinal and transverse polarization of the produced hadron. Superscript $\perp$ specifies the quark transverse momentum dependence FFs and subscript $1$ denotes the leading-twist FFs. One dimensional or integrated FFs can be obtained by taking integration over $k_T$. Therefore, only three one dimensional FFs are left, they are $D_1(z)$, $G_{1L}(z)$ and $H_{1T}(z)$.

Without the parity constraint, we need to consider the P-odd FFs which are induced by the non-trivial $\theta$-vacuum tunneling effects.
\begin{align}
  &z\Xi^P_\alpha=\bar n_\alpha\bigg[\mathcal{D}_1+\frac{\varepsilon_{T kS}}{M}\mathcal{D}^\perp_{1T}\bigg],\label{f:zxiPO}\\
  &z\tilde\Xi^P_\alpha=\bar n_\alpha\bigg[\lambda \mathcal{G}_{1L}+\frac{k_T \cdot S_T}{M}\mathcal{G}^\perp_{1T}\bigg],\label{f:zxi5PO}\\
  &z\Xi^P_{\rho\alpha}=\bar n_\rho\bigg[\frac{k_{T\alpha}}{M}\mathcal{H}^\perp_{1} +\varepsilon_{T S\alpha}\mathcal{H}_{1T}+\frac{\varepsilon_{T k\alpha}}{M}\mathcal{H}^\perp_{1S}\bigg].\label{f:zxioddPO}
\end{align}
where superscript $P$ is used to represent the P-odd quantities, $\mathcal{H}^\perp_{1S}=\lambda \mathcal{H}^\perp_{1L}+\frac{k_T \cdot S_T}{M}\mathcal{H}^\perp_{1T}$. $\mathcal{D}$, $\mathcal{G}$ and $\mathcal{H}$ are used to denote the unpolarized, longitudinal polarized and transverse polarized quarks fragmenting into hadrons. 
They have a one-to-one correspondence to $D, G$ and $H$. Hence, among the total 16 FFs, 8 of them are parity conserved, and the others are parity violated.
The one dimensional P-odd FFs are similar to the P-even ones, they are $\mathcal{D}_1(z)$, $\mathcal{G}_{1L}(z)$ and $\mathcal{H}_{1T}(z)$.

We note here that the definitions of the P-odd FFs given in Eqs. (\ref{f:zxiPO})-(\ref{f:zxi5PO}) are different from the definitions in ref. \cite{Yang:2019rrn}, Eqs. (2.15)-(2.16). The previous definitions are misleading, e.g., $D_{1T}^\perp$ corresponds to $\mathcal{G}_{1T}^\perp$ rather than $\mathcal{D}_{1T}^\perp$ in the former definitions. This would be misunderstood. To avoid this misunderstanding, we require that the P-odd FFs have the very same forms to the P-even FFs in this paper. The cost of these definitions is that P-odd factors should be introduced in the decomposition of the quark-quark correlator in order to keep the cross section P-even, e.g., a $\gamma^5$ factor should be introduced in Eq. (\ref{f:zxiPO}).

Leading-twist one dimensional FFs have an interpretation as probability densities. $D_1(z)$ is the number density of finding an unpolarized hadron inside an unpolarized quark. $G_{1L}$ is the number density difference of quarks with helicity $+$ and quarks with helicity $-$. It is known as  longitudinal spin transfer function. $H_{1T}(z)$ is the transverse spin transfer function which is interpreted as the number density difference of quarks with transverse polarization $\uparrow$ and quarks with transverse polarization $\downarrow$. We note that $H_{1T}$ admits the probabilistic interpretation only in the transverse polarization basis. For the P-odd FFs, they do not have the probabilistic interpretations, they are only reflections of the parity violations. However, the P-odd FFs are not the simple extensions of the P-even ones, they reflect the complex QCD vacuum structure. Furthermore, azimuthal asymmetries induced by the P-odd TMD FFs provide us an alternative to study the non-trivial QCD vacuum \cite{Yang:2019rrn}.

For the higher twist P-even and P-odd FFs, they can also be obtained by decomposing the quark-quark and quark-j-gluon-quark correlators \cite{Chen:2016moq,Yang:2017sxz}. We do not repeat the decompositions in this paper. We also note here that the way we introducing these FFs seems different from ref. \cite{Mulders:1995dh}. However, they can yield the same results. There is one more advantage to use employ the definition in this paper or ref. \cite{Chen:2016moq,Yang:2017sxz}. When higher twist contributions are taken into consideration, the relationships between the quark-quark correlators and quark-gluon-quark correlators obtained form the QCD equation of motion are often used to eliminate the independent FFs. By using the definitions used in ref. \cite{Chen:2016moq,Yang:2017sxz}, the relationships can be written as unified forms.

\section{Properties of fragmentation functions}\label{S:bounds}

\begin{figure}[t]

  \centering
  \includegraphics[width=5cm]{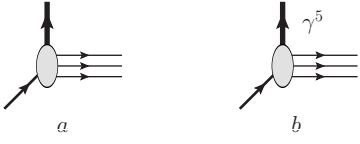}\\
  \caption{The P-even (a) and P-odd (b) parton-hadron vertices.}\label{Fig:vertices}
\end{figure}

By using the optical theorem, the fragmentation process can be described by the u-channel forward amplitude, $\mathcal{A}_{\lambda\Lambda'\lambda'\Lambda}$, where $\lambda, \lambda'$ are the helicities of the incoming and outgoing quarks while $\Lambda, \Lambda'$ are the helicities of the initial and final hadrons. There are 16 amplitudes in total, as a consequence of the parity, time-reversal and helicity conservation constraints, only three independent amplitudes are left, 
\begin{align}
  &\mathcal{A}_{++,++}, &&\mathcal{A}_{+-,+-}, &\mathcal{A}_{++,--}. \label{f:amplitudes}
\end{align}
To illustrate the positivity bounds, we first define the P-even and P-odd quark-hadron vertices $a_{\lambda\Lambda'}$ and $a^P_{\lambda\Lambda'}$, see Fig.~\ref{Fig:vertices}. We can use the optical theorem to relate the amplitudes to the three leading-twist one dimensional FFs \cite{Soffer:1994ww,Goldstein:1995ek,Metz:2016swz}:,
\begin{align}
  &D_1 \sim Im(\mathcal{A}_{++,++}+\mathcal{A}_{+-,+-}) \sim \sum_X(a^*_{++}a_{++}+a^*_{+-}a_{+-}), \label{f:D1Im}\\
  &G_{1L} \sim Im(\mathcal{A}_{++,++}-\mathcal{A}_{+-,+-}) \sim \sum_X(a^*_{++}a_{++}-a^*_{+-}a_{+-}), \label{f:G1LIm}\\
  &H_{1T} \sim Im\mathcal{A}_{++,--}\sim \sum_X a^*_{--}a_{++}. \label{f:H1TIm}
\end{align}
In the transversity basis $H_{1T}$ can be expressed as $Im(\mathcal{A}_{\uparrow\uparrow,\uparrow\uparrow}-\mathcal{A}_{\uparrow\downarrow,\uparrow\downarrow})$, where $\uparrow$ points in $y$ direction. From Eqs. (\ref{f:D1Im})-(\ref{f:H1TIm}), we can write down the following relations immediately,
\begin{align}
  &D_1(z) \geq |G_{1L}(z)|, 
  &&D_1(z) \geq |H_{1T}(z)|. \label{f:D1dayueH1T}
\end{align}
Considering the inequality relation, $\sum_X |a_{++} \pm a_{--}|^2 \geq 0$,
and using the parity invariance constraint, we can obtain the Soffer's inequality
\begin{align}
  D_1(z) + G_{1L}(z) \geq 2|H_{1T}(z)|. \label{f:Soffer}
\end{align}


The previous discussions of the P-even FFs are not novel. But, they can be applied to the P-odd case equally, hence we present them explicitly here. In the following context, we show a general discussion and introduce a definition of the P-odd correlator based on this discussion. We consider the P-odd and P-even FFs simultaneously. For the amplitudes shown above, we have
\begin{align}
  \mathcal{A}^t_{++,++} &\sim \sum_X\big(a^*_{++}+a^{P*}_{++}\big)\big(a_{++}+a^{P}_{++}\big) \nonumber\\
  &= \sum_X\big(2a^*_{++}a_{++}+ a^{P*}_{++}a_{++}+ a^*_{++}a^{P}_{++}\big), \label{f:Azzzz} \\
  \mathcal{A}^t_{+-,+-} &\sim \sum_X\big(a^*_{+-}+a^{P*}_{+-}\big)\big(a_{+-}+a^{P}_{+-}\big) \nonumber\\
  &= \sum_X\big(2a^*_{+-}a_{+-}+ a^{P*}_{+-}a_{+-}+ a^*_{+-}a^{P}_{+-}\big), \label{f:Azfzf} \\
  \mathcal{A}^t_{++,--} &\sim \sum_X\big(a^*_{--}+a^{P*}_{--}\big)\big(a_{++}+a^{P}_{++}\big) \nonumber\\
  &= \sum_X\big(2a^*_{--}a_{++}+ a^{P*}_{--}a_{++}+ a^*_{--}a^{P}_{++}\big), \label{f:Azzff}
\end{align}
where supercript $t$ denotes the whole amplitude including both the P-even vertex and the P-odd vertex. Those first terms on the right hand side in Eqs. (\ref{f:Azzzz})-(\ref{f:Azzff}) correspond to the P-even FFs while other terms contribute to the P-odd ones. To obtain these relations, we have used $a^*_{\lambda\Lambda'}a_{\lambda\Lambda'}= a^{P*}_{\lambda\Lambda'}a^{P}_{\lambda\Lambda'}$. For the P-odd amplitudes, we have $\mathcal{A}^P_{\lambda\Lambda',\lambda'\Lambda} \sim 2 Re (a^{P*}_{\lambda\Lambda'}a_{\lambda'\Lambda})$. Thus, the inequalities shown above also apply here:
\begin{align}
  &\mathcal{D}_1 \geq |\mathcal{G}_{1L}|, &&\mathcal{D}_1 \geq |\mathcal{H}_{1T}|, && \mathcal{D}_1 + \mathcal{G}_{1L}\geq 2|\mathcal{H}_{1T}|. \label{f:inequalityPodd}
\end{align}

Here we need an explanation to the P-odd amplitudes. There are 16 amplitudes in total, but only three of them are left when impose the parity, time-reversal and helicity conservation constraints. We note that the parity constraint used to eliminate the correlated amplitudes puts no constraint on the P-odd amplitudes (vertices). The reason is P-odd amplitudes are localized in a small domain, the parity violated effects can be reduced completely when sum over all the events at the macroscopic level. This means the P-odd effect in QCD is local and it can only be ``seen" on the event-on-event basis. In other words, the P-odd amplitudes are local quantities, they do not contradict to the global parity constraint.

The correlator shown in Eq. (\ref{f:correlationfunction}) contains both the P-even and P-odd components \cite{Kang:2010qx}. However, it is more convenient to separate the two components for more detailed discussions,
\begin{align}
  \hat \Xi(k,p_h)\to \hat\Xi'(k,p_h)=\hat\Xi(k,p_h)+ \hat\Xi^P(k,p_h). \label{f:Xiprime}
\end{align}
Inspired by the current theory ($j^\mu=\bar \psi\gamma^\mu\psi, j^{\mu5}=\bar \psi\gamma^\mu\gamma^5\psi$) and the weak interaction ($\bar \psi\gamma^\mu(c_V^f\pm c_A^f\gamma^5)\psi$), we can introduce the P-odd correlator immediately by the replacement, $\psi \to \gamma^5 \psi$. Furthermore, by introducing the P-odd parton-hadron vertex, we obtained the P-odd amplitudes. According to the previous discussion, it can be seen that the inequalities of the P-odd FFs are straightforward extensions of the P-even ones. This indicates the P-odd and P-even correlators should have similar forms and the replacement is reasonable. However, we should keep in mind that the replacement is just a formal one. In this case, we introduce the P-odd correlator with the following form,
\begin{align}
  \hat \Xi^P(k,p_h)= \int \frac{d^4\xi}{2\pi} & e^{ik\xi} \sum_X \Big[\langle 0 |\psi(\xi)|p_h, X\rangle \langle p_h, X|\bar\psi(0)\gamma^5|0 \rangle\nonumber\\
  + &\langle 0 |\gamma^5\psi(\xi)|p_h, X\rangle \langle p_h, X|\bar\psi(0)|0 \rangle \Big]. \label{f:correlationfunctionODD}
\end{align}
This definition may not be the unique definition of the P-odd correlator. However, it seems that it is an economical one and it can help us for more detailed discussions of the P-odd FFs. We will see in the following context.

Parity odd quantities induced by the non-trivial $\theta$ vacuum tunneling effect are local quantities, FFs are of no exception. This means that these FFs should vanish when sum over all the final hadrons h because of being local quantities. To study this unique property of these FFs, in the following context we give a proof to illustrate this.
For now we start from the definition of the P-even correlator. The P-odd one can be given in a similar way. The P-even correlator can be rewritten with the creation ($a_h^\dag$) and annihilation ($a_h$) operators \cite{Collins:1981uw},
\begin{align}
  \hat\Xi(k,p_h)=\frac{1}{2\pi} \int d^4 e^{ik\xi} \langle 0 |\psi(\xi) a_h^\dag a_h\bar\psi(0)|0 \rangle . \label{f:CollinsDefinition}
\end{align}

We note here that the FFs obtained above are defined in a reference frame where the produced hadron has no transverse momentum. For the following calculations, it is convenient to switch to a frame where the quark does not have the transverse momentum.
We first calculate the following integration,
\begin{align}
  &\int dz\cdot  z \cdot \hat\Xi^{[\Gamma]}(z)
  = \int dz \frac{d^2p_{h\perp}}{(2\pi)^2} z\cdot \hat\Xi^{[\Gamma]}(z,p_{h}) \label{f:dzzD1} \\
  =& \int \frac{d\xi^-d^2\xi_T}{2k^+} \frac{dp_h^+d^2p_{h\perp}}{(2\pi)^32p^+_{h}} e^{ik\xi} Tr \langle 0 |\psi(\xi)a_h^\dag p_h^+ a_h\bar\psi(0)\Gamma|0 \rangle\Big|_{\xi^+=0},\nonumber
\end{align}
where $\hat \Xi^{[\Gamma]}(z)$ denotes one dimensional FF specified by the $\Gamma$-matrix. By summing over all hadrons h, the integration can be rewritten as
\begin{align}
  &\sum_h\int dz\cdot z \cdot \hat\Xi^{[\Gamma]}(z) \nonumber\\
  &=\int \frac{d\xi^-d^2\xi_T}{2k^+} e^{ik\xi} Tr \langle 0 |\psi(\xi) p^+ \bar\psi(0)\Gamma|0 \rangle \Big|_{\xi^+=0}, \label{f:sumh}
\end{align}
where the momentum operator $p^+$ is given by
\begin{align}
  p^+ =\sum_h\int \frac{dp_h^+d^2p_{h\perp}}{(2\pi)^32p^+_{h}}a_h^\dag p^+_h a_h. \label{f:momentumoperator}
\end{align}
The sum of all hadrons h plays a important role. It indicates that contributions of all of the final hadrons h can be replaced by a single one when the momentum operator acts on a special state. So, we
insert a complete set of quark states and obtain the following expression
\begin{align}
  &\sum_h\int  dz\cdot  z \cdot \hat \Xi^{[\Gamma]}(z)= \int \frac{d\xi^-d^2\xi_T}{2k^+} e^{ik\xi} Tr \langle 0 |\psi(\xi)\nonumber\\
  &\times\sum_{s'}\int \frac{dk^{\prime+}d^2k'_T}{(2\pi)^32k^{\prime+}}|k',s' \rangle p^+ \langle k',s'|\bar\psi(0)\Gamma |0 \rangle\Big|_{\xi^+=0}, \label{f:quarkset}
\end{align}
where $k', s'$ denotes the quark momentum and spin, respectively. By using $\psi(x)|k\rangle =u(k)e^{-ikx}|0\rangle$ and taking the integration over $k$ and $\xi$, we finally have
\begin{align}
  &\sum_h\int dz\cdot z \cdot \hat\Xi^{[\Gamma]}(z) =\frac{1}{4k^+}Tr[u(k)\bar u(k) \Gamma]. \label{f:result}
\end{align}
If $\Gamma=\gamma^+$, we can obtain $\sum_{h}\int dz\cdot z \cdot D_1(z) =1/2$ with $\bar u(k)\gamma^+u(k)=2k^+$.
As argued in ref. \cite{Meissner:2010cc}, $D_1$ is defined by a spin average rather than a spin summation for a spin 1/2 hadron. Hence, one has to multiply a factor $(2s+1)$ for a spin 1/2 hadron to obtain the sum rule of $D_1(z)$,
\begin{align}
  \sum_{h}\sum_S \int dz\cdot z \cdot D_1(z) =1.\label{f:sumrule}
\end{align}
The sum of the hadron spins is important and this is the main reason why the sum rules of the polarization dependence FFs do not exist.
As mentioned before, the sum of all hadrons h is crucial in the derivation. It indicates that the sum of all the hadrons' momenta is equal to the quark's momentum. In other words, the momentum sun rule, Eq. (\ref{f:sumrule}), is equivalent to the momentum conservation law in fragmentation process. All the hadrons h act as a single quark.

Following the same derivation given in Eqs. (\ref{f:dzzD1})-(\ref{f:result}) and using the definition of the P-odd correlator, we can obtain the following equation
\begin{align}
  &\sum_h\int dz\cdot z \cdot \mathcal{D}(z) =0, \label{f:P-oddD1}
\end{align}
where we have used $\bar u(k)\gamma^5\gamma^+u(k)=-\bar u(k)\gamma^+\gamma^5 u(k)$. Equation (\ref{f:P-oddD1}) indicate P-odd FF $\mathcal{D}(z)$ has no contribution to the physical measurement when sum over all the hadrons h, in other words $\mathcal{D}(z)$ is a local quantity and can be ``seen" only on the event-by-event basis. If $\Gamma$ is taken as $\gamma^+\gamma^5$, we can obtain the similar equation for $ \mathcal{G}_{1L}(z) $
\begin{align}
  &\sum_h\int dz\cdot z \cdot \mathcal{G}_{1L}(z) =0. \label{f:P-oddG1L}
\end{align}
The detailed derivations are shown in appendix.
It is straightforward to obtain the relations of $\mathcal{D}(z)$ and $\mathcal{G}_{1L}(z)$. However, $\mathcal{H}_{1T}(z)$ it is another thing. Because $\mathcal{H}_{1T}(z)$ is chiral-odd FF, it must company with another chiral-odd FF to contribute to the cross section.
In this case, we consider the following way.

As we know that the differential cross section can be expressed as a contraction of the leptonic tensor and hadronic tensor in high energy reactions. In annihilation process the hadronic tensor can be written as a trace
\begin{align}
  W^{\mu\nu}= Tr\left[\gamma^\mu \hat\Xi(z) \gamma^\nu \bar{\hat\Xi}(z)\right]. \label{f:hadronictensor}
\end{align}
Decomposing the correlators in Eq. (\ref{f:hadronictensor}) with only chiral-odd FFs at leading twist, the hadronic tensor can be given by the following trace
\begin{align}
   W^{\mu\nu}=-Tr\left[\gamma^\mu \gamma^i \slashed{\bar{n}} \gamma^\nu \gamma^j \slashed n \right]S^i_{1T}S^j_{2T}{H}_{1T}(z)\bar{H}_{1T}(z). \label{f:hadronictrace}
\end{align}
We see that Eq. (\ref{f:hadronictrace}) is just the term which rises the double spin asymmetry in annihilation process. For the calculation of the P-odd process, we recall the replacement  $\psi \to \gamma^5 \psi$ by introducing the definition of the correlator. Since there is one more factor $\gamma^5$ in the correlator, the decomposition of the correlator with Dirac matrices must also have one more $\gamma^5$. This just corresponds to the statement at the end of the second paragraph below Eq. (\ref{f:zxioddPO}) and proves the importance of introducing the definition of P-odd correlator. As mentioned at the end of the third paragraph below Eq. (\ref{f:zxioddPO}), the P-odd FFs are not the simple extensions of the P-even ones, but corollaries of the decomposition of the P-odd correlator (Eq. (\ref{f:correlationfunctionODD})). By introducing a $\gamma^5$ in the trace, we have
\begin{align}
W^{P\mu\nu}\sim Tr\left[\gamma^\mu \gamma^i \slashed{\bar{n}} \gamma^\nu \gamma^j \slashed n \gamma^5 \right]S^i_{1T}S^j_{2T}\mathcal{H}_{1T}(z)\bar{\mathcal{H}}_{1T}(z). \label{f:hadronictraceOdd}
\end{align}
It is can be easily checked that Eq. (\ref{f:hadronictraceOdd}) gives no contribution, as expected, when contracts with the leptonic tensor. Leptons are not affected by the strong interactions, so the leptonic tensor is unchanged. As mentioned before, the sum rule of $D_1(z)$ denotes the momentum conservation law in fragmentation process. However, Eq. (\ref{f:P-oddD1}) does not indicate the violation of momentum conservation law. It denotes that P-odd FF $\mathcal{D}_1(z)$ is a local quantity and vanish when sum over all hadrons h. Because the sum of all the hadrons can completely reduce the non-trivial $\theta$-vacuum tunneling effects by averaging all the fluctuation events. The same arguments apply to $\mathcal{G}_{1L}(z)$ and $\mathcal{H}_{1T}(z)$.

Even though the P-odd FFs vanish when sum over all the hadrons h because of local quantities, they can be measured on the event-by-event basis. By applying the notations used in this paper, it is can be shown that the double spin asymmetry has one more origin \cite{Yang:2019rrn},
e.g., $A^{\cos(\phi_{S1}+\phi_{S2})}_{TT}\sim H_{1T}\bar H_{1T}+\mathcal{H}_{1T}\bar{\mathcal{H}}_{1T}$. From the quarks point of view, it corresponds to the quark-antiquark transverse polarization correlation $c_{nn}^q$, see Fig. \ref{Fig:cnny} (The derivation is given in the appendix.).
If quarks propagate in the domain where magnetic field $B$ acts as background field, the quarks' spins can be aligned along the $B$ field direction. It means that the quark-antiquark transverse polarization correlation $c_{nn}^q$ is violated. It is can be seen from the deformations of the curves shown in Fig. \ref{Fig:cnny}.
Though the alternative origin of the double spin asymmetry that companies with $H_{1T}$ makes it difficult to extract the $\cos(\phi_{S1}+\phi_{S2})$ effect from experiments measurement, it is important to determine the P-odd FFs.
\begin{figure}[t]
  \centering
  \includegraphics[width=6cm]{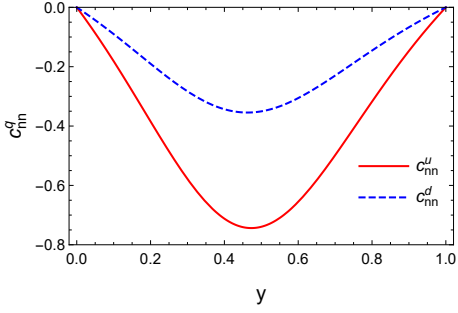}\\
  \caption{The quark-antiquark transverse polarization correlation at $Z^0$ pole, $Q=M_Z$. The horizontal coordinate $y$ is a function of scattering angle $\theta$ in c.m.s., $y=(1+\cos\theta)/2$.}\label{Fig:cnny}
\end{figure}

\section{summary}\label{S:summary}

When quarks enter in P-odd domains formed by the non-trivial $\theta$-vacuum tunneling effects, the P-odd FFs can be induced in fragmentation processes. In this paper we give a discussion about these P-odd FFs.
In the quantum field theoretical formulation, FFs are given by the quark-quark correlator. Since correlator is a $4\times 4$ matrix in Dirac space, we can decompose them with the Dirac matrices. By decomposing the correlator, we obtain 16 FFs in total at leading twist, 8 of the are parity violated while the others are parity conserved. These P-odd FFs and P-even FFs have a one-to-one correspondence and they can be discussed simultaneously. To study the positivity bounds, we define both the P-even and P-odd parton-hadron vertices. We find that the inequalities of the P-even one dimensional FFs still apply to the P-odd ones. In our discussions, we find that it is convenient to separate the P-odd and the P-even components in the correlator, hence we have $\hat\Xi'(k,p_h)=\hat\Xi(k,p_h)+ \hat\Xi^P(k,p_h)$. Inspired by the current theory ($j^\mu=\bar \psi\gamma^\mu\psi, j^{\mu5}=\bar \psi\gamma^\mu\gamma^5\psi$) and the weak interaction ($\bar \psi\gamma^\mu(c_V^f\pm c_A^f\gamma^5)\psi$), we introduce an economical definition of the P-odd correlator. We emphasize again the importance of introducing the definition of the P-odd correlator. It not only provides us an economical way to study the properties of P-odd FFs but also indicates that the P-odd FFs are not the simple extensions of the P-even ones but corollaries of the decomposition of the P-odd correlator.
Based on this definition we present a proof that P-odd FFs vanish when sum over all hadrons h because of being local quantities.
An argument is that the sum of all the hadrons can completely reduce the non-trivial $\theta$-vacuum tunneling effects by averaging all the fluctuation events. However, local quantities can be measured on the event-by-event basis.  For example, double spin asymmetry, $A^{\cos(\phi_{S1}+\phi_{S2})}_{TT}\sim H_{1T}\bar H_{1T}+\mathcal{H}_{1T}\bar{\mathcal{H}}_{1T}$, can be used to extract the P-odd FFs on this basis.

When the TMD FFs are taken into consideration, the sensitive quantities studied in experiments are often different azimuthal asymmetries. By measuring these asymmetries, we can extract the P-odd FFs which are important quantities in revealing the non-trivial vacuum structures in QCD in high energy reactions. However, the measurement of electric dipole moment of neutron indicates a stringent limits on $\theta$ ($<3\times 10^{-10}$). This may disappoint us because of the small magnitude of these asymmetries. But this can also inspire us to study more about the dynamics of the interactions in the P-odd domains. Furthermore, measurements of the P-odd asymmetries provide us an alternative way to study the parity violation in QCD. Parton distribution functions are taken as the counterparts of FFs, they shall have the same properties.



\begin{appendix}

\section{Derivation of the vanishing local quantities}

The steps to calculate Eq. (\ref{f:P-oddD1}) and (\ref{f:P-oddG1L}) are similar to Eq. (\ref{f:sumrule}).
Following the same steps, we first calculate the following integration,
\begin{align}
  &\int dz\cdot  z \cdot \hat\Xi_1^{[\Gamma]}(z)
  = \int dz \frac{d^2p_{h\perp}}{(2\pi)^2} z\cdot \hat \Xi_1^{[\Gamma]}(z,p_{h}) \label{f:PdzzD1} \\
  =& \int \frac{d\xi^-d^2\xi_T}{2k^+} \frac{dp_h^+d^2p_{h\perp}}{(2\pi)^32p^+_{h}} e^{ik\xi} Tr \langle 0 |\psi(\xi)a_h^\dag p_h^+ a_h\bar\psi(0)\gamma^5\Gamma|0 \rangle\Big|_{\xi^+=0}. \nonumber
\end{align}
Here we only consider the first part in the definition of the P-odd correlator, the second part can be shown in the same way. By summing over all hadrons h and using the definition of the momentum operator, one finds
\begin{align}
  &\sum_h\int dz\cdot z \cdot \hat\Xi_1^{[\Gamma]}(z) \nonumber\\
  &=\int \frac{d\xi^-d^2\xi_T}{2k^+} e^{ik\xi} Tr \langle 0 |\psi(\xi) p^+ \bar\psi(0)\gamma^5\Gamma|0 \rangle \Big|_{\xi^+=0}, \label{f:sumhD1}
\end{align}
Inserting a complete set of quark states and using $\psi(x)|k\rangle =u(k)e^{-ikx}|0\rangle$, we have
\begin{align}
  &\sum_h\int  dz\cdot  z \cdot \hat\Xi_1^{[\Gamma]}(z)= \int \frac{d\xi^-d^2\xi_T}{2k^+} e^{ik\xi} Tr \langle 0 |\psi(\xi)\nonumber\\
  &\times\sum_{s'}\int \frac{dk^{\prime+}d^2k'_T}{(2\pi)^32k^{\prime+}}|k',s' \rangle p^+ \langle k',s'|\bar\psi(0)\gamma^5\Gamma |0 \rangle\Big|_{\xi^+=0} \nonumber\\
  & =\frac{1}{4k^+}Tr[u(k)\bar u(k) \gamma^5 \Gamma ]. \label{f:firstD1}
\end{align}
When the second part in the P-odd correlator is added, we obtain the complete equation
\begin{align}
  &\sum_h\int  dz\cdot  z \cdot \hat\Xi^{[\Gamma]}(z) \nonumber \\
  & =\frac{1}{4k^+}\Big(Tr[u(k)\bar u(k) \gamma^5 \Gamma ]+Tr[u(k)\bar u(k)\Gamma \gamma^5 ]\Big). \label{f:tildeXi}
\end{align}
By inserting $\Gamma=\gamma^+, \gamma^+\gamma^5$, we can obtain
\begin{align}
  &\sum_h\int dz\cdot z \cdot \mathcal{D}(z) =0,
  &\sum_h\int dz\cdot z \cdot \mathcal{G}_{1L}(z) =0. \label{f:D1G1L}
\end{align}

To be explicit, we require the first part in the P-odd correlator has ``positive" contribution.  Then the sum of all the hadrons h act as a single ``positive" quark while the counterpart or second part acts as a single ``negative" quark with the same contribution. Assuming there is a function $f^P$ denoting the implicit contribution, thus we have $f_p^P(k)+f_n^P(k)=0$. (We note here $f^P$ is only introduced to interpret Eq. (\ref{f:D1G1L}). Further discussions are beyond the scope of this paper.)

\section{Quark-antiquark transverse polarization correlation}

When both electromagnetic and weak interactions are taken into consideration, the differential cross section of the electron-positron annihilation process $(e^+ e^-\to q\bar q)$ can be given by
\begin{align}
  \frac{d\sigma}{dy} &=\frac{2N_c \pi \alpha^2}{s}\Big\{\chi T^q_0(y) + \chi^q_{int} T_V^q(y)+e_q^2A(y)\Big\}, \label{eq:ch2-eeqq}
\end{align}
where $y=(1+\cos\theta)/2$, $\theta$ is the scattering angle, $A(y)=(1-y)^2+y^2=(1+\cos^2\theta)/2$, $s=Q^2$. The coefficients $N_c$, $\alpha$ and $e_q$ denote the color factor, fine structure constant and quark charge, respectively. $T_0^q(y)$ and $T_V^q(y)$ are given by
\begin{align}
  T^q_0(y)&=c_1^ec_1^qA(y)-c_3^ec_3^qB(y), \label{eq:ch2-T0}\\
  T^q_V(y)&=c_V^ec_V^qA(y)-c_A^ec_A^qB(y). \label{eq:ch2-T1}
\end{align}
where $B(y)=1-2y=-\cos\theta$. $c_V^{e,q}$ and $c_A^{e,q}$ are weak coupling constants which satisfy $c_1^i=(c_V^i)^2+(c_A^i)^2, c_3^i=2c_v^ic_A^i$, $i=e, q$. The coefficients~$\chi$ and $\chi_{int}^q$ are
\begin{align}
&\chi=\frac{s^2}{[(s-M_Z^2)^2+\Gamma_Z^2 M_Z^2]\sin^42\theta_W},\\
&\chi_{int}^q = \frac{- 2 e_q s (s - M_Z^2)}{ [(s-M_Z^2)^2 + \Gamma_Z^2 M_Z^2] \sin^2 2\theta_W},
\end{align}
where $ M_Z$ and $\Gamma_Z$ are the mass and decay width of $Z^0$ boson. $\theta_W$ is Weinberg angle.

In electron-positron annihilation process, the quark-antiquark transverse polarization correlation is defined as \cite{Chen:2016iey}
\begin{align}
 c_{nn}^q\equiv\frac{|\hat m_{n++}|^2+|\hat m _{n--}|^2-|\hat m_{n+-}|^2-|\hat m_{n-+}|^2}{|\hat m_{n++}|^2+|\hat m_{n--}|^2+|\hat m_{n+-}|^2+|\hat m_{n-+}|^2}, \label{f:cnn}
\end{align}
where $\hat m$ is the helicity amplitudes and the subscripts $+$ or $-$ denotes quark or antiquark is in $s_n=\frac{1}{2}$ or $ s_n=-\frac{1}{2}$ states. Here $\vec n$ is taken as the normal of the the production plane. By calculating the helicity amplitudes, the quark-antiquark transverse polarization correlation can be calculate as
\begin{align}
  c^q_{nn}(y, Q)=-\frac{C(y)\big[\chi c_1^ec_2^q+\chi_{int}^q c_V^ec_V^q+e_q^2\big]}{2\big[\chi T_0^q(y) + \chi^q_{int} T_V^q(y)+e_q^2A(y)\big]}s_{1T}\cdot s_{2T}, \label{f:Qcorrelation}
\end{align}
where $c_2^q = (c_V^q)^2-(c_A^q)^2$, $C(y)=4y(1-y)=\sin^2\theta$. $s_{1T}$ and $s_{2T}$ denote the quark and antiquark transverse polarizations and satisfy $s_{1T}\cdot s_{2T}=-|s_{1T}||s_{2T}|$. For simplicity, hereafter we only consider the weak interaction in the following derivation. In this case, we can obtain the following equation,
\begin{align}
 c_{nn}^{q}(y)=-\frac{c_1^ec_2^qC(y)}{2T_0^q(y)}s_{1T}\cdot s_{2T}.  \label{f:cnnqy}
\end{align}
To see the angle dependence of the quark-antiquark transverse polarization correlation, we plot Eq. (\ref{f:cnnqy}) in Fig. \ref{Fig:cnny}.

Under the condition of the collinear factorization, we obtain
\begin{align}
 c_{nn}^{h_1,h_2}=|s_{1T}||s_{2T}|\frac{c_1^ec_2^qC(y)}{2T_0^q(y)}\frac{ D_1(z_1)\bar D_1(z_2)}{D_1(z_1)\bar D_1(z_2)}  \label{f:cnnh1h2}
\end{align}
by multiplying $D_1(z_1)\bar D_1(z_2)$ in both the numerator and denominator. By utilizing $s_T D_1(z)=S_T H_{1T}(z)$ \cite{Barone:2001sp}, we have
\begin{align}
 c_{nn}^{h_1,h_2}=|S_{1T}||S_{2T}|\frac{c_1^ec_2^qC(y)}{2T_0^q(y)}\frac{ H_{1T}(z_1)\bar H_{1T}(z_2)}{D_1(z_1)\bar D_1(z_2)}.  \label{f:CH1H2}
\end{align}
In our derivation we ignore the interchange between the quark and antiquark FFs because it does not affect the derivation. One may find out that we require $\vec n$ being the normal of the the production plane to obtain Eq. (\ref{f:CH1H2}). If this constraint is released, that is $\vec n$ could point anywhere, we have
\begin{align}
  \tilde{c}_{nn}^{h_1,h_2}=|S_{1T}||S_{2T}|\frac{c_1^ec_2^qC(y)\cos(\phi_{s_1}+\phi_{s_2})}{2T_0^q(y)}\frac{H_{1T}(z_1)\bar H_{1T}(z_2)}{D_1(z_1)\bar D_1(z_2)}. \label{eq:ch2-CDSA}
\end{align}
We can see that Eq. (\ref{eq:ch2-CDSA}) just corresponds to the double spin asymmetry.

\end{appendix}

\end{document}